# Structured light by discrete-phase orbital angular momentum holograms

A. VIJAYAKUMAR,[1] CARMELO ROSALES-GUZMAN,[2,3,*] MANI RATNAM RAI,[1] JOSEPH ROSEN,[1] OLEG V. MININ,[4] IGOR V. MININ,[5] AND ANDREW FORBES[2]

[1]*Department of Electrical and Computer Engineering, Ben-Gurion University of the Negev, P.O. Box 653, Beer-Sheva 8410501, Israel*
[2]*School of Physics, University of the Witwatersrand, Private Bag X5, Johannesburg 2050, South Africa*
[3]*Wang Da-Heng Collaborative Innovation Center for Quantum manipulation & Control, Harbin University of Science and Technology, Harbin 150080, China.*
[4]*National Research Tomsk State University, Lenin Avenue, 36, Tomsk, 634050, Russia*
[5]*National Research Tomsk Polytechnic University, Lenin Avenue, 30, Tomsk, 634050, Russia*
*\* carmelorosalesg@gmail.com*

**Abstract:** Structured light has been created by a myriad of near- and far-field techniques and has found both classical and quantum applications. In the case of orbital angular momentum (OAM), continuous spiral phase patterns in dynamic or geometric phase are often employed with the phase patterns existing across the entire transverse plane. Here we exploit the uncertainty relation between OAM and angle to create structured OAM fields using multilevel OAM holograms. We show theoretically and experimentally that only a multilevel angular phase contour in the near-field is needed to create structured OAM light in the far-field, exploiting the reciprocal nature of angular momentum and angle. We use this approach to demonstrate exotic 3D structured light control to show the evolution of the Poynting vector in such fields and to highlight the physics underlying this phenomenon.

## 1. Introduction

Structured light concerns the tailoring of light in amplitude, phase and polarization, and has become highly topical lately [1, 2]. Particular attention has been paid to modes that carry orbital angular momentum (OAM) [3,4] which have found many applications and have been extensively reviewed to date (see Ref. [5] and references therein). Such beams are characterized by an azimuthal phase dependency $exp(im\phi)$, where $m$ is the topological charge, $\phi$ is the azimuthal angle, and carry $m\hbar$ of OAM per photon. Because of the interest in utilizing OAM for optical communication as quantum [6], classical [7] and hybrid states [8], there has been considerable attention placed on the generation and detection of such modes. There are several options available for the detection of OAM modes, from fork holograms in reverse [9], diffraction off apertures [10, 11], and mode transformation optics [12]. For a quantitative measure that includes intermodal phases, the modal decomposition approach has been shown to be highly versatile [13, 14]. Generation approaches include dynamic phase in the form of spiral phases, usually encoded on spatial light modulators [14, 15], geometric phase in the form of liquid crystals [16, 17] or meta-surfaces [18], non-spiral phase plates [19] and a suite of approaches for intra-cavity generation [20]. Coupled to this has been recent interest in tailoring OAM fields in 3D for the demonstration of exotic propagation characteristics [21–31] as well as encoding fractional OAM on the field [32].

In these studies, the desired azimuthal phase is usually encoded across the entire transverse plane. But we point out that for the helicity of light to be well-defined requires only a well-defined contour in the plane that encloses the required azimuthal phase gradient, and not a gradient everywhere. Further, the OAM structure of the field may be engineered by employing the tools of diffractive optics, binarising the phase gradient into a multilevel element. We show that this allows us to engineer OAM spectrums of fields, from single valued to a desired spread, and explain this concept in the language of Fourier optics applied to angular diffraction [33]. We test the concept experimentally by quantitatively measuring the OAM spectrum, finding excellent agreement between theory and experiment. Finally, we use this approach to design simple holograms for the control of complex 3D structured light based on OAM. Since multilevel (binary) holograms are cheap and easy to fabricate, this approach should be useful in real-world applications of OAM.

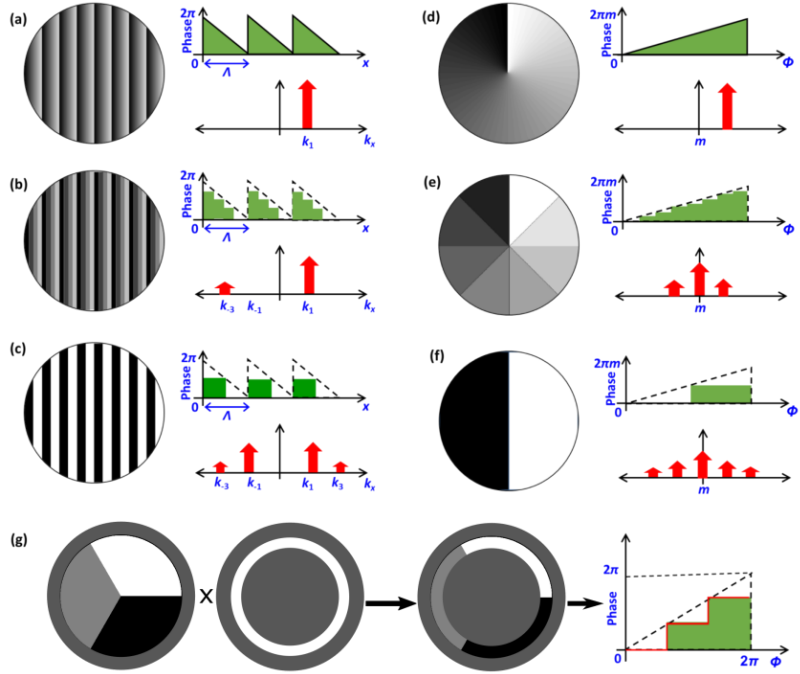

Fig. 1. A schematic of the concept. Converting a continuous blazed grating (a) into a multilevel element results in a spectrum of linear momenta, shown in (b) and (c). Likewise, a continuous azimuthal phase in (d) has a well-defined OAM value, whereas a multilevel approximation in the azimuth, (e) and (f), results in a spectrum of OAM. In (g) the multilevel approach is confined to only a contour in the transverse plane, allowing for the simple generation of controlled OAM superpositions.

## 2.  Multilevel OAM: concept and theory

The early days of diffractive optical elements (DOEs) saw lithography processes utilized for the creation of DOEs and hence it was (and still is) commonplace to approximate the kinoform with a multilevel element comprising a binary step-like function, with the consequence of several diffraction orders but the advantage of tailored efficiency, as illustrated in Fig. 1 (a) - (c). One can think of this as an uncertainty relation: becoming more certain in linear position (say a one-step approximation to the function) means less certainty in linear momentum, hence several wavevectors produce the diffraction orders. Just as linear position and linear momentum share a reciprocal uncertainty relation (Heisenberg's uncertainty principle), so the same is true for angular position (angle) and angular momentum [33]. If we now consider a binary approximation to the angles, then as we become more certain in angle, so we become less certain in OAM, shown in Fig. 1 (d) - (f). This translates into a well-defined OAM spectrum that is a function of the multilevel approximation used in the segmentation of the angle. Further, we note that the topological charge of the mode is specified by a contour integral enclosing the desired azimuthal phase, given by

$$m = \frac{1}{2\pi} \oint \nabla \Phi(x,y) \cdot d\vec{s}, \tag{1}$$

where $\Phi$ is the phase profile and $m$ the resulting topological charge. This allows, for example, OAM Bessel modes to be created by defining OAM in an annular ring while ignoring the rest of the transverse plane [34–36], as illustrated in Fig. 1 (g). We put forward that this implies the possibility to engineer OAM modes by multilevel binary holograms that segment the angle along a particular contour. Thus, the *type* of OAM mode will depend on what the radial segmentation looks like (how the transverse plane is partitioned), e.g., Bessel for a ring slit, but the OAM spectrum itself is independent of this. Since multilevel binary holograms are cheap and easy to fabricate, this approach should be useful in real-world applications of OAM.

To illustrate this mathematically, consider the generic case of an incoming Gaussian beam, $u_0(r) = exp\left[-\left(\frac{r}{w_0}\right)^2\right]$, where $w_0$ is the beam waist, incident on a $N$-level hologram with a transmission function

$$t(\phi) = \sum_{n=1}^{N} exp(i\phi_n) \quad with \quad \frac{2\pi m(n-1)}{N} \leq \phi \leq \frac{2\pi mn}{N}. \tag{2}$$

This defines the angular segments, e.g., $\phi_1 \in \left[0, \frac{2\pi m}{N}\right]$, but not the values within the segments, which can be arbitrarily chosen. For illustrative purposes we will select a typical multilevel approximation so that each segment is valued at the average value of the range (see Fig. 2), namely,

$$\phi_n = \frac{2\pi m(n-1)}{N}. \tag{3}$$

If we now wish to study only the OAM content of such an output field, $U(r,\phi) = u_0(r)t(\phi)$, we can integrate out the radial term and express the remaining azimuthal dependence in the angular basis

$$U(\phi) = \sum_m A_m \, exp(im\phi), \tag{4}$$

where the modal coefficients are given by

$$A_m = \frac{1}{2\pi} \int_0^{2\pi} U(\phi) \, exp(-im\phi) \, d\phi. \tag{5}$$

Note that the basis is orthonormal and complete so that the spectrum satisfies $\sum_m |A_m|^2 = 1$, where $|A_m|^2$ is the power of the $m^{th}$ OAM mode. It is easy to show that for our example choice of multilevel OAM hologram one finds for the specific case of an input beam $m_{in}=1$

$$A_m = \frac{exp\left(\frac{i\pi m}{N}\right)}{N} \, sinc\left(\frac{\pi m}{N}\right) \frac{sin(\pi m)}{sin\left[\frac{\pi(m-1)}{N}\right]}. \tag{6}$$

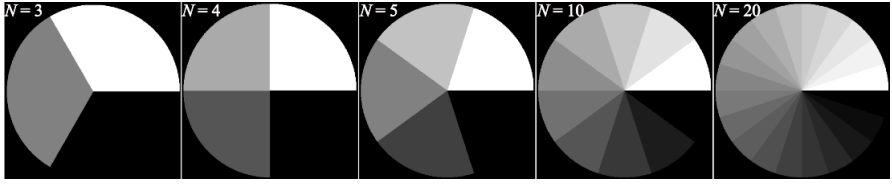

Fig. 2. Example of angular partitioning the transverse plane to create multilevel binary holograms. As the number of phase levels (N) is increased so the azimuthal phase becomes closer to the ideal, in this example a topological charge of $m = 1$. The black color indicates a 0-phase value, while white indicates a $2\pi$-phase value.

With the "to be designed" field now specified in terms of the decomposition in Eqs. (4) and (5), we can easily find the physical properties of the field such as the Poynting vector

$$S = \frac{\varepsilon_0 \omega}{4} [i(U\nabla U^* - U^*\nabla U) + 2k|U|^2 \hat{z}], \quad (7)$$

where $\varepsilon_0$ is the free space permittivity and $k$ is the wavenumber. The orbital angular momentum density in the $z$ direction given by

$$j_z = r \times \frac{S}{c^2}, \quad (8)$$

the total angular momentum $\mathbf{J_z}$ per unit length is given by

$$J_z = \iint_\mathbb{R} j_z \cdot dA, \quad (9)$$

and the effective topological charge of the beam, inferred from the ratio between its total OAM $\mathbf{J_z}$ and total energy $W$ per unit length is given by

$$\frac{J_z}{W} = \frac{\iint_\mathbb{R} (r \times \langle E \times B \rangle_z) \cdot dA}{c \iint_\mathbb{R} \langle E \times B \rangle_z \cdot dA} = \frac{m_{eff}}{\omega}. \quad (10)$$

Here $\omega$ is the angular frequency of the wave and $c$ is the speed of light in vacuum. All of these physical properties can easily be expressed in terms of the spectrum coefficients, $A_m$.

## 3. Experiments and Results

### 3.1 Experimental set-up

The experimental set-up is shown in Fig. 3. Light from a He-Ne laser ($\lambda = 632.8$ nm) was spatially filtered using a pinhole and collimated using a lens $L_1$ with a focal length of $f_1 = 200$ mm. The collimated light was polarized using a polarizer P along the active axis of a phase-only reflective spatial light modulator ($SLM_1$) (Holoeye PLUTO, 1920 × 1080 pixels, 8 μm pixel pitch, phase-only modulation). In the $SLM_1$, a phase mask obtained by a product of $t(\phi)$ of Eq. (2) and linear phase given by $L(\bar{v}) = exp\{i2\pi(\lambda)^{-1}(v_x x + v_y y)\}$ was displayed for different number of phase levels (the levels in our multilevel element). The resulting phase pattern (fork grating) [37], was used to modulate the incident light. The light modulated by $SLM_1$ was passed through a relay system made up of two identical lenses $L_2$ and $L_3$ each with a focal length of $f_2 = 100$ mm. An aperture $A_1$ located at the common focal plane filtered out only the first diffraction order and blocked the higher orders. This optical configuration was used to project the complex amplitude generated after the $SLM_1$ onto a second SLM ($SLM_2$) (identical to $SLM_1$). In $SLM_2$, a hologram obtained by spatially multiplexing holograms corresponding to different azimuthal modes was displayed [15,16,37--] to simultaneously detect the entire OAM spectrum by modal decomposition. In the experiment, a total of 20 holograms were multiplexed while the inset in the Fig. 3 shows the hologram obtained by multiplexing of three holograms. The light modulated by the SLM$_2$ was Fourier transformed using lens L$_3$ with a focal length of f$_3$ = 150 mm on the image sensor 1 (Flea3 from Pointgrey, 1280×1024 pixels, 5.3 μm pixelsize). An aperture A$_2$ was used at the entrance of the image sensor to remove the higher diffraction orders. An array of diffraction patterns was produced, and bright spots were obtained for the cases when the phase of the incident beam matched with the azimuthal mode of the multiplexed hologram. Therefore, by observing the diffraction patterns, the modal composition of the light beam could be determined accurately [11-13]. The first experiment is denoted as channel – C$_1$.

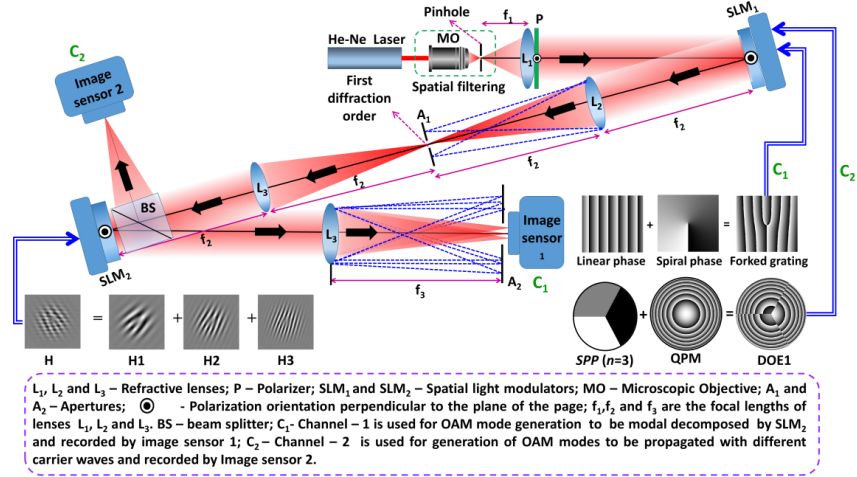

Fig. 3. Experimental set-up used for generation of beams by a multilevel phase element as well as the subsequent modal decomposition of the beams.

A second channel C$_2$ was created by inserting a beam splitter (BS) at the entrance of the second SLM (SLM$_2$), which redirects the beam towards another image sensor (Thorlabs DCC3240M CMOS, pixel pitch = 5.3 μm, 1280 × 1024 pixels). On the SLM$_1$, a phase pattern obtained by combining a multilevel spiral phase plate (*SPP*) with a carrier wave generating diffractive optical function was displayed [38]. Therefore, the same experimental set up can be used for the study of modal decomposition and for subsequent study of 3D beam propagation for a particular multilevel phase approximation.

### 3.2 Binary OAM spectrums

The OAM spectrum of the generated light field was determined experimentally by optical means using modal decomposition [14-16]. For this, we projected the input field into a set of orthogonal OAM modes and measure the on-axis intensity in the far field. To this end, a second SLM is used, where the transmission functions $exp(-im\phi)$ are displayed in the form of a fork hologram. Crucially, even though in theory the number of required modes is infinite, experimentally only a finite subset is necessary. In our experiments we restricted the topological charge of the input mode to $m_{in} \leq |4|$, which only required the subset of modes $m \in [-10,9]$ for a full decomposition. Needless to say, for higher topological charges we need larger subsets of projecting modes. The full spectrum was determined in a single shot by multiplexing all the transmission functions into a single hologram, as illustrated in the inset of Fig. 3 (SLM$_2$). Moreover, a calibration hologram allows the identification of the on-axis coordinates for each multiplexed mode, as shown in Fig. 4 (a). An example of the far field intensity pattern of an $m_{in} = 4$ input mode projected onto the multiplexed subset of modes is shown in Fig. 4 (b), a bright on-axis spot at $m = 4$ discloses its topological charge.

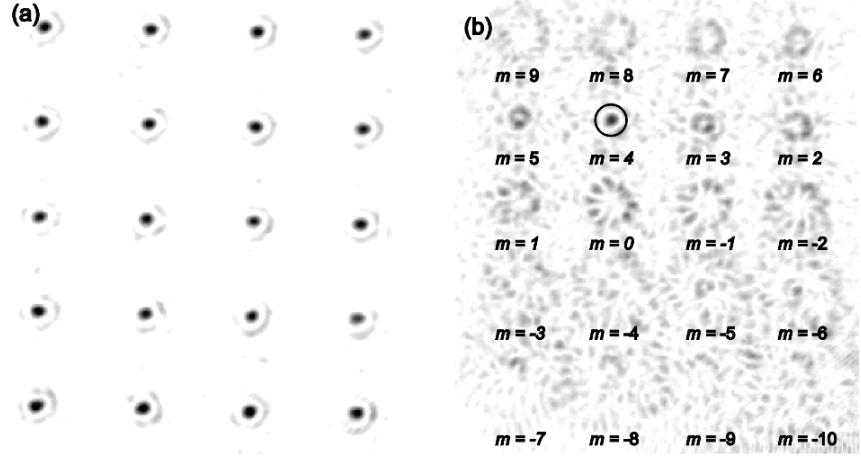

Fig. 4. Far field intensity of multiplexed modes. (a) Calibration pattern. (b) Projection of an $m_{in}$=4 input mode onto the multiplexed set of modes $exp(-im\phi)$, $m \in [-10,9]$.

Figure 5 shows the OAM spectrum of an input field ($m_{in}$=3) as function of the number of phase steps ($N$) employed in its generation, simulation in (a) and experiment in (b). Importantly, the modal coefficients $|A_m|^2$, $m \neq 3$, converges to zero with increasing values of $N$, whereas $|A_3|^2$ converges to one. For this particular example, only 15 phase steps ($N$=15) are required to generate a highly pure OAM mode.

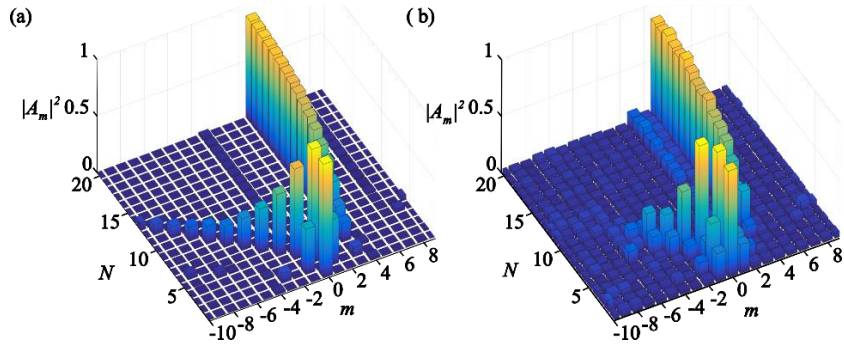

Fig. 5. Simulation (a) and experimental (b) measurement of the OAM spectrum of an input beam ($m_{in}$=3) as a function of the phase steps ($N$) employed for its generation.

As can be expected, the minimum number of phase steps required to generate an OAM mode, depends on the topological charge of the same. To illustrate this, we measured the OAM spectrum for the cases $m_{in} = \pm 4$ [Fig. 6(a) and 6(d)], $m_{in} = \pm 2$ [Fig. 6(b) and (e)] and $m_{in} = \pm 1$ [Fig. 6(c) and 6(f)] as function of $N$. Here for example, to generate an OAM mode with topological charge $m_{in} = \pm 4$ we require at least 15 steps. Importantly, as will be shown later, to generate an OAM mode with topological charge $m_{in} = \pm 1$ (or at least a very good approximation to it), only 3 phase steps are sufficient [see Fig 6 (c) and 6(f)]. Notice also that the OAM spectrum features a non-symmetric behavior. The OAM spectrum for input modes with topological charge $m_{in} < 0$ only contains OAM values to the right of the given mode, analogously, for $m_{in} > 0$ the OAM spectrum is formed by modes on the left side.

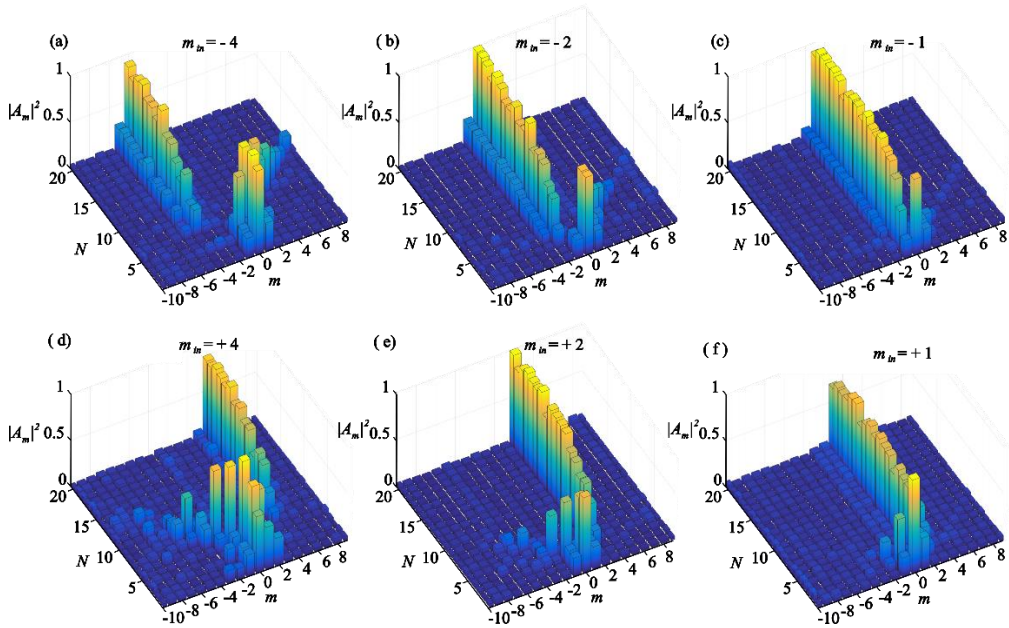

Fig. 6. Experimental measurements of the OAM spectrum as a function of the number of phase steps (*N*) employed to generate an input beam with topological charge $m_{in}$.

## *3.3 Energy flow and OAM in 3D*

In the previous section we showed experimentally that in order to generate an OAM mode with topological charge $\boldsymbol{m_{in} = \pm 1}$, only three phase levels (*N* = 3) are required. The case of two phase levels (0, π) results from a coherent superposition of two beams with OAM values *m* = 1 and *m* = −1 and as a result, the total OAM is zero. Therefore, with two phase levels, the OAM spectrum will always have a large *m* = 0 component, regardless of the input beam. For three phase levels, this symmetry is broken, and becomes a better approximation to the topological charge of the input beam. To see this, notice that from Eq. (6) the coefficients $|\boldsymbol{A_m}|^2$, related to the power content of each mode, can be written as,

$$|A_m|^2 = \begin{cases} \frac{N^2}{(\pi m)^2}, & \frac{m-1}{N} \in \mathbb{Z} \\ 0, & \frac{m-1}{N} \notin \mathbb{Z} \end{cases}. \tag{11}$$

From this equation we can see that $|\boldsymbol{A_m}|^2$ increases with $N^2$ and decreases with $m^2$. In other words, for a fixed value of *N*, the coefficients $|\boldsymbol{A_m}|^2$ decrease very rapidly as *m* increases. In particular, for the special case *N* = 3, $|\boldsymbol{A_m}|^2$ takes non-zero values at *m* = […, -5, -2, 1, 4, …], namely, $|\boldsymbol{A_m}|^2 = \frac{9}{\pi^2}[\dots, \frac{1}{25}, \frac{1}{4}, \boldsymbol{1}, \frac{1}{16}, \dots]$. That is, for an input mode with topological charge *m* = −1, with only three phase levels the coefficient $|\boldsymbol{A_1}|^2$ is relatively high compared to the rest. In the following section, this case will be further analyzed from an application point of view.

In what follows, we will consider the generation of more complicated phase structures by combining our *N*=3 *SPP* with other phase functions, namely, a quadratic phase mask (QPM) [39], an axilens [40] and an accelerating phase function [41]. Further, we analyzed the evolution of the Poynting vector of the generated beams as they propagate in free space. The experimental set-up for this is shown in Fig. 3, the generating holograms were displayed in SLM$_1$ (see bottom-right insets).

We start with the combination of a three-level *SPP* ($\Phi_{SPP}$) and a *QPM*, given by $\Phi_{QPM}$= [-$(\pi/\lambda f)r^2]_{2\pi}$, where $r^2=(x^2+y^2)$ and $f$ = 500 mm in order to bring the far-field intensity pattern of SPP at the focal plane of the *QPM*. The modulo-$2\pi$ phase addition of *SPP* for $m = 1$ and $N = 3$ with $\Phi_{QPM}$ results in a diffractive optical element (DOE) given by $\Phi_{DOE1}=[\Phi_{SPP}+\Phi_{QPM}]_{2\pi}$ as shown in Fig. 3. The collimated, polarized light was modulated by the phase pattern displayed on the SLM and the intensity pattern was recorded by an image sensor-2 located at a distance of about 500 mm from the SLM. Assuming a uniform illumination of the SLM, the image of the simulated intensity pattern obtained by $\left|exp(i\Phi_{DOE}) * exp\left(\frac{i\pi r^2}{\lambda f}\right)\right|^2$, where '*' is a 2D convolution operator is shown in Fig. 7 [42]. The Poynting vector, which in the case of scalar optical fields can be calculated as the product of the gradient of the phase and the simulated intensity distribution of the beam, is plotted on Fig. 7 (a) [43]. While the intensity pattern of the beam generated experimentally and recorded by the image sensor is shown in Fig. 7 (b). From Fig. 7 (a), the presence of orbital angular momentum (OAM) is seen with only three phase levels. The comparison of Fig. 7 (a) and Fig. 7 (b) shows a good match between the experimental and simulated results.

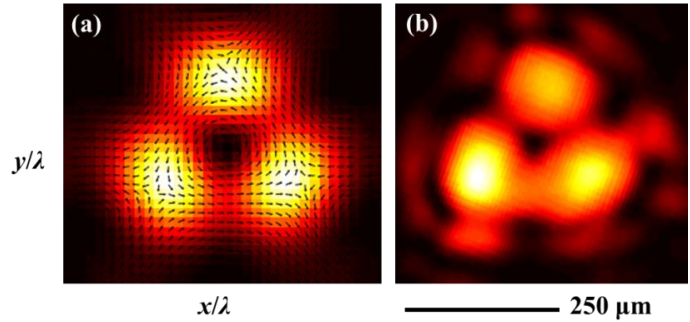

Fig. 7. (a) Simulated intensity at the focal plane of the diffractive lens using Fresnel propagation and Poynting vector plot, (b) experimental intensity recorded at the focal plane.

As a second example, we synthesized a Bessel-like beam using an axilens, with a phase function given by $\Phi_{AL}$= [-$(\pi/\lambda f_r)r^2]_{2\pi}$. Here $f_r=f_0+\Delta f(r^2/R^2)$, $f_0$ being the focal length of the axilens, with $\Delta f$ the focal depth and $R$ the radius of the element. The axilens was designed for $f_0$= 48 *cm*, $\Delta f$=10 *cm* and *R*=8 *mm* [41,45]. The phase function of the axilens was combined with the $m = 1$, *N*=3 *SPP*, using modulo-$2\pi$ phase addition $\Phi_{DOE2}=[\Phi_{SPP}+\Phi_{AL}]_{2\pi}$. The phase image of the DOE1 and the experimental and simulated results of the 2D intensity patterns for three different planes (*z*=45 *cm*, 48 *cm* and 51 *cm*) are indicated in Fig. 8 (a) and (c). The propagation of the intensity and the Poynting vector fields in the 3D space is given in visualization 1. From the longitudinal intensity pattern and the Poynting vector variation shown in visualization 1, the flow of the optical field indicates that the OAM is preserved as the beam propagates in free space.

Finally, we consider the combination of our SPP with an accelerating phase function. First, the propagation trajectory of the airy beam can be tailored using a curved phase function given by [41,45] $\Phi_C = \left[C\frac{2\pi}{\lambda}\left(\sqrt{x-x_0}+\sqrt{y-y_0}\right)\right]_{2\pi}$, where *C* is a constant. The length of the path over which the intensity pattern must be preserved is controlled using an axilens as described earlier. The phase pattern modulo-$2\pi$ resulting from the addition of the curved phase function, the *SPP* and the axilens $\Phi_{DOE3}=[\Phi_{SPP}+\Phi_{AL}+\Phi_C]_{2\pi}$ is shown in Fig. 8 (b). The 2D intensity patterns simulated using beam propagation and recorded in the experiment for three different planes (*z*=45 *cm*, 48 *cm* and 51 *cm*) are shown in Fig. 8 (d). The propagation of the optical field with the Poynting vector plots are given in visualization 2. The phase function $\Phi_C$ generates a carrier wave which propagates the intensity and Poynting vector fields in a curved path. From visualization 2, even in the curved path, the intensity profile and the Poynting vector fields were maintained. There was a slight redistribution of intensity among the three spots in some regions of the path of the beam, however, the Poynting vector plots indicate the presence of OAM throughout the path of the carrier waves. In the absence of $\Phi_C$, there was no redistribution of light among the three spots.

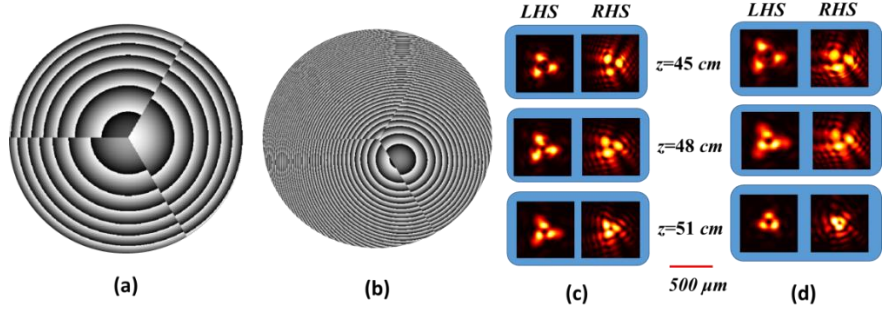

Fig. 8. Phase image of (a) DOE1 and (b) DOE2. The simulated (LHS) and experimental (RHS) results at three different planes ($z$=45 $cm$, 48 $cm$ and 51 $cm$) for (c) DOE1 (visualization 1) and (b) DOE2 (visualization 2).

## 4. Discussion and Conclusion

In this study, the modal decomposition method has been implemented to understand accurately the number of phase levels required to generate an OAM mode accurately. There is no doubt that with a higher number of phase levels the accuracy with which an OAM mode can be generated increases. However, from a manufacturing point of view, it is often challenging to manufacture phase elements with higher number of phase levels besides the cost and time involved in creating the multiple phase masks and the fabrication procedure [46]. Therefore, lesser the number of phase levels, easier is the fabrication procedure. From this new study, a standard can be reached to decide on the number of phase levels necessary for generating an OAM mode accurately. For a general phase element such as a grating or a Fresnel zone plate, it is well-known from the previous studies that the efficiency can be traded off with a much simpler fabrication procedure involving only two-phase levels [46]. From this study, it is seen that a symmetric phase configuration involving two phase levels is not suitable for generation of a fundamental OAM mode with $m = 1$ and requires at least three phase levels ($n = 3$) to obtain a Poynting vector map demonstrating the presence of OAM.

Besides, the multifunctional DOEs designed by modulo-$2\pi$ phase addition of the three level *SPP* and carrier waves generating phase functions shows that the OAM can be retained over a longer range in 3D space. From the recent fabrication attempts and challenges seen in the fabrication of *SPP* with higher number of phase levels, it is more convenient to reduce the fabrication complexities based on the recent findings to minimum number of phase levels [47].

**Funding**

The work of AV, MRR and JR was supported by the Israel Science Foundation (ISF) (Grant No. 1669/16) and by the Israel Ministry of Science and Technology (MOST).

**Acknowledgement**

This study was done during a Research Stay of JR at the Alfried Krupp Wissenschaftskolleg Greifswald. AV thanks Dr. Vijay Kumar, IISER Pune, India for the useful discussions.